\documentclass[11pt]{article}
\usepackage{amsmath}
\usepackage{amssymb}

\usepackage{graphicx}
\usepackage{bbm,bm}  
\usepackage{cite}
\usepackage[numbers,sort&compress]{natbib}

\usepackage[colorlinks,
  linkcolor=blue,
  anchorcolor=black,
  citecolor=blue
]{hyperref}

\usepackage{caption}

\usepackage{subcaption}

\begin{document}

\def\bfej{\mbox{\boldmath$\varepsilon$}}
\newcommand{\ccbar}{c\bar c}
\newcommand {\sla}[1]{ #1 \!\!\!/}
\newcommand{\jgmm}{\jp\to \gamm MM}
\newcommand{\jp}{J/\psi}
\newcommand{\st}[1]{|#1\rangle}
\newcommand{\jgpp}{\jp\to\gamma\phi\phi}
\def \qqbar {q\bar q}
\newcommand{\str}[1]{|#1\rangle}
\newcommand{\elmnt}[3]{\langle #1|#2|#3\rangle}
\newcommand{\cp}{CP~}
\newcommand{\cpv}{CP violation~}

\begin{center}

\par
{\Large\bf  Perturbative QCD Prediction of  the Hyperon EDM  from CP-violating Dipole 
Interactions } 
\par\vskip20pt
Kai-Bao Chen$^1$,  Xiao-Gang He$^{2,3}$,  Jian-Ping Ma$^{4,5}$ and Xuan-Bo Tong$^{6,7}$    \\
{\small {\it
$^1$ School of Science, Shandong Jianzhu University, Jinan, Shandong 250101, China\\
$^2$ State Key Laboratory of Dark Matter Physics, Tsung-Dao Lee Institute, Shanghai Jiao Tong University, Shanghai 201210, China\\
$^3$ Shanghai Key Laboratory for Particle Physics and Cosmology,
Key Laboratory for Particle Astrophysics and Cosmology (MOE),
School of Physics and Astronomy, Shanghai Jiao Tong University, Shanghai 201210, China\\
$^4$ School of Physics, Henan Normal University, Xinxiang, Henan 453007,  China\\
$^5$ Institute of Theoretical Physics, P.O. Box 2735, Chinese Academy of Sciences, Beijing 100190, China\\
$^6$ Department of Physics, University of Jyväskylä, P.O. Box 35, 40014 University of Jyväskylä, Finland\\
$^7$ Helsinki Institute of Physics, P.O. Box 64, 00014 University of Helsinki, Finland\\}} 

\end{center}

\abstract{Electric dipole moment (EDM) of baryons provides a sensitive probe of CP-violating interactions beyond the Standard Model. Motivated by the recent BESIII measurement on the $\Lambda$-hyperon EDM~\cite{BESIII:2025vxm}, we present the first perturbative QCD analysis of the $\Lambda$ EDM form factor to elucidate its origin in CP-violating quark dipole interactions.  In particular, we derive a QCD factorization formula that relates the $\Lambda$ EDM form factor to quark EDMs and chromo-electric dipole moments (CEDMs) through convolutions with light-cone distribution amplitudes of $\Lambda$. These connections allow us to extract constraints on CP-violating dipole couplings from current and future hyperon EDM measurements. Our numerical analysis demonstrates that the $\Lambda$ EDM exhibits unique sensitivity to the strange-quark CEDM, providing complementary information to that obtained from the neutron EDM.}

\par
\vskip20pt
\noindent

It is well known that the Standard Model~(SM) has CP-violating interactions characterized through a nonzero phase in the CKM matrix elements. This phase successfully accounts for the observed CP violation in meson systems at high-energy experiments. However, the SM phase alone is insufficient to explain the large baryon–antibaryon asymmetry of the Universe when combined with the standard cosmological model. This limitation has motivated the long-term pursuit of new sources of CP violation beyond the SM and their experimental exploration.

The existence of a nonzero electric dipole moment (EDM) of a particle is a clear signal of CP violation. Since EDMs predicted by the SM, arising from the CKM phase, are extremely small,  the existence of an EDM would imply new CP-violating interactions beyond the SM. While the EDMs of the electron and neutron have been extensively studied both theoretically and experimentally (see~\cite{Alarcon:2022ero,Chupp:2017rkp} and references therein), experimental investigations of hyperon EDMs are scarce. In fact, prior to the recent measurement by the BESIII collaboration~\cite{BESIII:2025vxm}, the only upper bound on the $\Lambda$ hyperon EDM had been established more than 40 years ago~\cite{FLD}.

In general,  a large number of events is required in high energy experiments to probe CP-violating effects or to test CP symmetry.  It has been proposed in~\cite{HMM,HeMa} to study the $\Lambda$ EDM $d_\Lambda$ by using large data samples of the $J/\psi \to \Lambda \bar\Lambda$ decays collected with BESIII detector. Recently, the BESIII collaboration has analyzed this decay and obtained a new upper bound on $d_\Lambda$~\cite{BESIII:2025vxm}:
\begin{equation}
   -8.6\times 10^{-19} <  {\rm Re }(d_\Lambda ) < 3.3\times 10^{-19}~e{\rm~cm}~, \quad   -2.5\times 10^{-19} <  {\rm Im }(d_\Lambda) < 7.2\times 10^{-19}~e{\rm~cm}~.
 \label{EQ1}
\end{equation} 
This result represents a significant improvement of nearly three orders of magnitude over the previous limit in~\cite{FLD}. With more accumulated data or at the planned Super Tau-Charm Factory~\cite{Achasov:2023gey}, further improvements are expected.

With the new result of $d_\Lambda$, an important question arises: how do CP-violating interactions among elementary particles like quarks generate a nonzero $d_\Lambda$?  The answer is essential for constructing new physics models beyond the SM. In this letter,
we consider the possibility that quarks possess a nonzero EDM and a nonzero Chromo-Electric Dipole Moment~(CEDM), and study their contributions to $d_\Lambda$. These dipole moments are introduced through a CP-violating effective Lagrangian as:
 \begin{eqnarray}
{\mathcal L}_{CP} =\sum_{q=u,d,s} \biggr ( -\frac{i}{2} d_q\bar q \gamma_5\sigma_{\mu\nu} q  F^{\mu\nu}-\frac{i}{2} \tilde d_q\bar q \gamma_5\sigma_{\mu\nu}G^{\mu\nu}  q \biggr ) ~,  
\label{LEDMs}
\end{eqnarray} 
where $d_q$ and $\tilde d_q$ denote the quark EDM and CEDM, respectively. $q$ is the quark field of $u$, $d$ and $s$. $F^{\mu\nu}$ or $G^{\mu\nu}$ is the field strength tensor of photon field or gluon field, respectively.  CP-violation is indicated by the existence of the EDM form factor defined as: 
\begin{equation}
   \langle 0 \vert J^\mu \vert \Lambda (P,\lambda)  \bar \Lambda (\bar P, \bar \lambda) \rangle = d_\Lambda(Q) \bar V(\bar P,\bar \lambda) \gamma_5 \sigma^{\mu\nu} q_\nu U(P,\lambda) + \cdots~, 
\label{DL}    
\end{equation} 
where $J^\mu$ is the electromagnetic current. $U$ and $\bar V$ are the spinor for $\Lambda$ and $\bar\Lambda$, respectively.  $Q$ is the invariant mass of the $\Lambda\bar\Lambda$ system. The $\cdots$ 
represent CP-conserved contributions. It should be emphasized that the defined  $d_\Lambda(Q)$ is not exactly the static EDM of $\Lambda$. It depends on $Q$. The static EDM corresponds to the value at $Q=0$, whereas the BESIII result in fact corresponds to $d_\Lambda(Q)$ evaluated at $Q = M_{J/\psi}$.

Using QCD low-energy effective field theories (EFTs) or lattice QCD, the contributions of the quark EDMs and CEDMs to the neutron EDM have been extensively studied~\cite{Alarcon:2022ero,Chupp:2017rkp}. These studies concern the static neutron EDM, i.e., evaluated at $Q=0$. In contrast, the $\Lambda$ EDM $d_\Lambda$ is extracted from the decay discussed above at an energy scale much larger than the nonperturbative scale $\Lambda_{\rm QCD}$. As a result, it is difficult to determine the contributions of $d_q$ or $\tilde d_q$ to $d_\Lambda$ at large $Q$ using EFTs or even quark models. To our knowledge, there is no lattice study of $d_\Lambda$ currently. In addition, we note that in the quark model, $d_\Lambda$ receives a contribution only from $d_s$, with $d_\Lambda = d_s$ at $Q=0$~\cite{DHMD}.

Inspired by the recent BESIII measurement~\cite{BESIII:2025vxm}, we present the first perturbative QCD prediction of the $\Lambda$ EDM form factor $d_\Lambda(Q)$ in the high-energy limit $Q \gg \Lambda_{\mathrm{QCD}}$ and derive the contributions from quark EDMs $d_q$ and CEDMs $\tilde d_q$. Our analysis is based on the collinear factorization framework for hard exclusive processes~\cite{Avdeenko:1981twg,Lepage:1979za,Brodsky:1981kj,Efremov:1979qk,Chernyak:1977as,Chernyak:1980dj,Lepage:1980fj,Chernyak:1983ej,Chernyak:1984bm}, which has been widely applied to various hadronic form factors~(see e.g.~\cite{Sun:2021gmi,Sun:2021pyw,Belitsky:2002kj,Tong:2021ctu,Tong:2022zax,Chen:2024fhj,Huang:2024ugd,Chen:2023byr,Ji:2024iak}). Within this framework, $d_\Lambda(Q)$ is expressed as a convolution of perturbatively calculable coefficient functions with the Light-Cone Distribution Amplitudes (LCDAs) of $\Lambda$, which encode its non-perturbative partonic structure. Using these formulas, we perform numerical estimates and investigate the constraints on the quark EDMs and CEDMs from the BESIII result. Our analysis shows that the $\Lambda$ EDM offers a unique opportunity to probe the strange-quark CEDM, in contrast to the neutron EDM.

We start by introducing the LCDAs of $\Lambda$. For our purpose, it suffices to compute $d_\Lambda(Q)$ at leading power in $1/Q$. At this order, only the leading twist or twist-3 LCDAs are needed, where the lowest Fock component of $\Lambda$, consisting of $s$,$u$ and $d$ quarks, is involved. We work in a light-cone coordinate system, in which a
vector $a^\mu$ is expressed as $a^\mu = (a^+, a^-, \vec a_\perp)$ with $a^{\pm}=(a^0\pm a^3)/\sqrt{2}$. Two light cone vectors $n^\mu = (0,1,0,0)$ and  $l^\mu =(1,0,0,0)$ in the system are introduced.  For a $\Lambda$ moving in the $z$-direction with the momentum $P$,   the leading twist or twist-3 LCDAs are defined as \cite{Braun:1999te}:
\begin{eqnarray}
&& 4\epsilon^{abc} \int \frac{ d\lambda_1 d\lambda_2 d\lambda_3}{(2\pi)^3} e^{ i P^+ (\lambda_1 x_1 + \lambda_2 x_2 + \lambda_3 x_3)}   \langle 0\vert u^a_{\alpha} (\lambda_3 n) d_{\beta}^b (\lambda_2 n) s^c_{\gamma} (\lambda_1n) \vert \Lambda(P)\rangle  
\nonumber\\
&&= {\mathcal V}_\Lambda ([x])  (\gamma\cdot p C)_{\alpha\beta} (\gamma_5 U)_\gamma 
      + {\mathcal A}_\Lambda ([x]) (\gamma\cdot p\gamma_5  C)_{\alpha\beta} U_\gamma 
  + {\mathcal T}_\Lambda ([x]) (p^\nu i\sigma_{\mu\nu} C)_{\alpha\beta}  (\gamma^\mu \gamma_5 U)  _\gamma +\cdots, 
\label{T3} 
\end{eqnarray} 
where $p^\mu =(P^+, 0,0,0)$ and $\cdots$ stands for higher-twist terms. The Latin letters denote color indices and the Greek ones denote the Dirac indices. $U$ is the spinor of $\Lambda$ with 
the momentum $p$, i.e., the mass of $\Lambda$ is neglected. ${\mathcal V}_\Lambda$, ${\mathcal A}_\Lambda$, and ${\mathcal T}_\Lambda$ are LCDAs of $\Lambda$. They depend on $x_{1,2,3}$ as the momentum fraction of corresponding parton, or quark. The dependence is denoted collectively as $[x]$. 
There is a constraint from the fact that $\Lambda$ is an isospin-0 baryon. This gives:
\begin{equation}
   {\mathcal V}_\Lambda  (x_1, x_2, x_3) =-  {\mathcal V}_\Lambda  (x_1, x_3,x_2)~,\quad  {\mathcal A}_\Lambda  (x_1,x_2,x_3) =  {\mathcal A}_\Lambda  (x_1,x_3,x_2)~,\quad    
    {\mathcal T}_\Lambda  (x_1, x_2, x_3) =-  {\mathcal T}_\Lambda  (x_1, x_3,x_2)~.  
\label{PRLCDA}     
\end{equation}    
The physical meaning of the LCDAs can be found if we express the state of $\Lambda$ in helicity basis~\cite{COZ}:
 \begin{eqnarray}
 \vert \Lambda^\uparrow\rangle &=& \int \frac{ [d ^3x ]}{4\sqrt{6}}  \biggr \{ \biggr [ {\mathcal V}_\Lambda ([x]) - {\mathcal A}_\Lambda ([x]) \biggr ] \vert u^\uparrow d^\downarrow s^\uparrow \rangle 
   + \biggr [ {\mathcal V}_\Lambda ([x] ) +  {\mathcal A}_\Lambda ([x]) \biggr ] \vert u^\downarrow d^\uparrow s^\uparrow \rangle  
-2{\mathcal T}_\Lambda ([x] ) \vert u^\uparrow d^\uparrow s^\downarrow \rangle\biggr \}~, 
\label{T3S} 
\end{eqnarray} 
with the notation $ [d^3 x] =d x_1 dx_2 dx_3 \delta (1-x_1-x_2-x_3)$. The upper index $\uparrow$ or $\downarrow$ denotes the $+$- or $-$-helicity, respectively. Through charge-conjugation transformation of Eq.(\ref{T3}) or Eq.(\ref{T3S}), one can obtain LCDAs of $\bar\Lambda$, denoted as ${\mathcal V}_{\bar\Lambda}$, ${\mathcal A}_{\bar\Lambda}$ and ${\mathcal T}_{\bar\Lambda}$.  
     
In the high energy limit, the mass of $\Lambda$ is neglected. The Dirac structure in Eq.~(\ref{DL}) flips the $\Lambda$ helicity, so the amplitude vanishes for $\lambda=\bar\lambda$ due to helicity conservation of massless fermions. Thus, only matrix elements with $\lambda\neq\bar\lambda$ need to be considered. Calculations of such helicity-flip matrix elements have their own interest in QCD. While helicity-conserving matrix elements related to the Dirac form factors have been studied (e.g., the recent one-loop calculations of nucleon Dirac form factor $F_1$~\cite{Chen:2024fhj,Huang:2024ugd}), there are only few results for helicity-flip form factors, even at tree level. Known examples include the nucleon Pauli form factor $F_2$, first calculated in~\cite{Belitsky:2002kj}, and certain gravitational form factors~\cite{Tong:2022zax,Tong:2021ctu}. In these cases, because QCD interactions conserve quark helicity, orbital angular momentum effects must be included, requiring twist-4 LCDAs to generate nonzero contributions. For $F_2 (Q)$, this gives an asymptotic scaling $F_2(Q)\sim Q^{-6}$. In contrast, the dipole interactions in Eq.~(\ref{LEDMs}) flip quark helicities directly, and thus flip the helicity of the $\Lambda$ relative to the $\bar\Lambda$. As a consequence, $d_\Lambda(Q)$ receives nonzero contributions already from twist-3 LCDAs, leading to the scaling behavior $d_\Lambda(Q)\sim Q^{-4}$. This scaling agrees with the general power-counting rules for exclusive processes~\cite{BF,Ji:2003fw}, and our explicit calculation confirms it.

\begin{figure}[t!!]
	\begin{center}
		\includegraphics[scale=0.37]{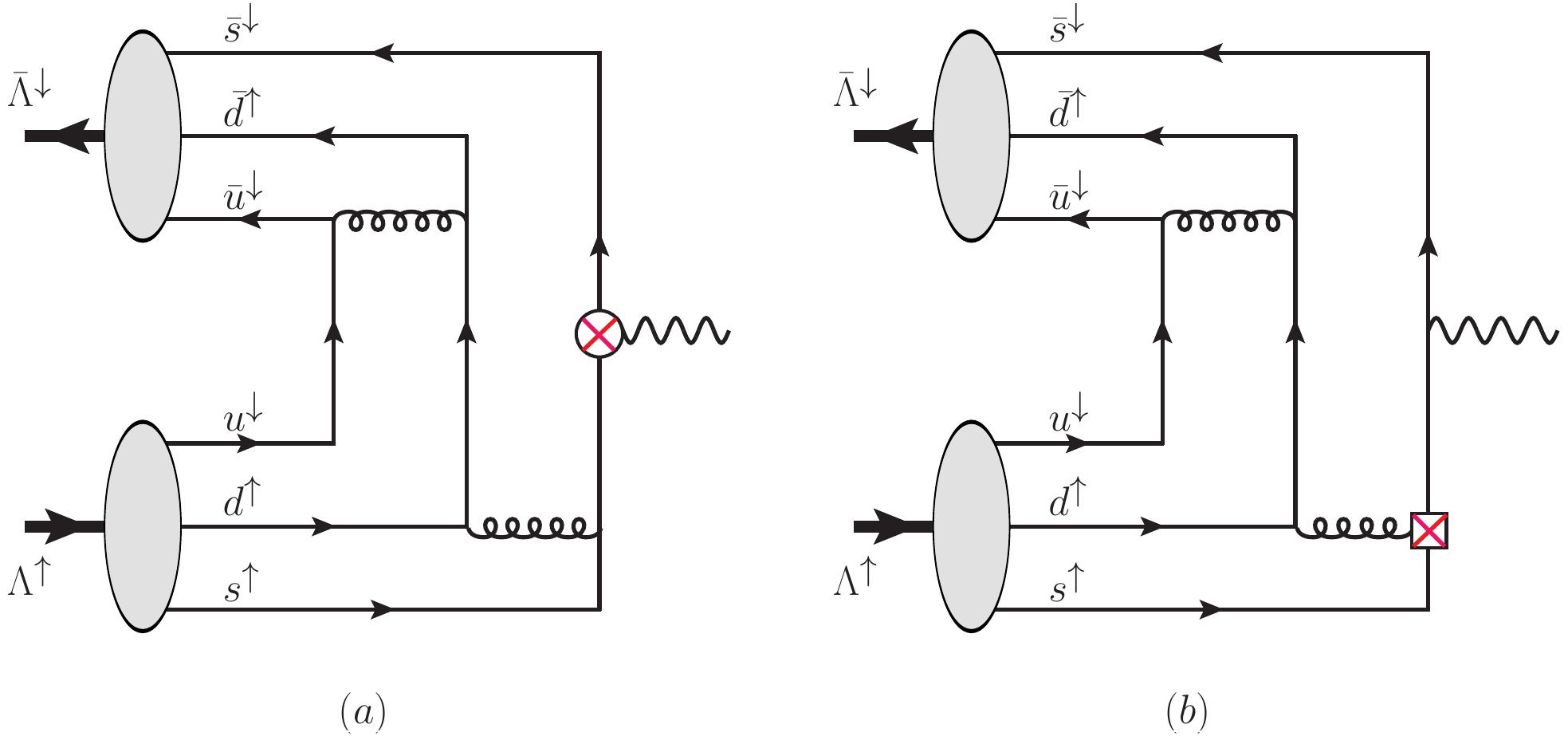}
  \end{center}
    \captionsetup{width=0.9\linewidth, font=small, labelfont=small} 
	\caption{(a). A typical diagram for the contribution from the EDM  $d_s$.  The photon vertex with a cross-circle 
	 is the EDM vertex. (b). A typical diagram for the contribution from the CEDM $\tilde d_s$.  The gluon vertex with a crossed square represents the CEDM vertex.   }
	\label{FF1}
\end{figure}

We proceed first to calculate the contribution from EDMs of quarks. A typical diagram for a contribution from $d_s$ is shown in Fig.~\ref{FF1}a. At large $Q$, hard gluon exchange between any pair of quark lines is required, and the different possible exchanges lead to a total of 14 diagrams. Because the three quarks are in color-singlet state, the contributions involving  three-gluon vertex are zero.  Letting the virtual photon line attach to the $u$- quark or $d$-quark, we obtain the contribution 
from $d_u$ or $d_d$, respectively.  There are 14 diagrams for the contributions from $d_u$ or $d_d$.  
Calculating these 42 diagrams, we derive the following contributions from EDMs of quarks: 
\begin{eqnarray} 
   d_\Lambda(Q)\biggr\vert_{\rm EDMs} &=&  C_B \frac{(4\pi \alpha_s)^2}{Q^4} \int [d^3x] [d^3y]
       \biggr \{  d_s \biggr [  \left ( {\mathcal V}_\Lambda ([x] ) {\mathcal V}_{\bar\Lambda} ([y]) -  {\mathcal A}_\Lambda ([x] ) {\mathcal A}_{\bar\Lambda} ([y]) \right )  {\mathcal H}_s ([x], [y] ) 
\nonumber\\
  && \quad + \left ( {\mathcal V}_\Lambda ([x] ) {\mathcal A}_{\bar\Lambda} ([y]) -  {\mathcal A}_\Lambda ([x] ) {\mathcal V}_{\bar\Lambda} ([y]) \right )  {\mathcal C}_s ([x], [y] ) \biggr ] 
\nonumber\\
          && +  
           (d_d+d_u)  \left ( {\mathcal V}_\Lambda ([x] ) + {\mathcal A}_\Lambda ([x] ) \right )  {\mathcal T}_{\bar\Lambda} ([y] )  {\mathcal H}_q ([x], [y] )   \biggr \}~,
\label{DEDMs} 
\end{eqnarray} 
with the perturbative coefficient functions at the leading order of $\alpha_s$:
\begin{eqnarray} 
  {\mathcal H}_s ([x], [y] )  &=& \frac{-2}{\bar x_1^2 \bar y_1^2  x_3 y_3}  - \frac{1}{2  x_2 x_3 y_2 y_3 }  \left ( \frac{1}{\bar x_3 \bar y_2} 
    -\frac{1}{\bar x_1 \bar y_2} -\frac{1}{\bar x_2 \bar y_1} \right ), 
\nonumber\\
    {\mathcal C}_s ([x], [y] ) &=& \frac{-1}{2 x_2 x_3 y_2 y_3} 
     \biggr ( \frac{1}{\bar x_3 \bar y_2} -\frac{1}{\bar x_1\bar y_2} + \frac{1} {\bar x_2 \bar y_1}\biggr ),
\nonumber\\
    {\mathcal H}_q([x],[y]) &=& \frac{-1}{ x_1 x_3 y_1 y_3} \left ( \frac{1}{\bar x_2 \bar y_3} -
    \frac{1}{\bar x_1 \bar y_3} +\frac{1}{\bar x_1 \bar y_2} \right ) + \frac{2}{ \bar x_2^2 \bar y_2^2} \left ( \frac{1}{x_3 y_3} + \frac{1} {x_1 y_1} \right ),                
\end{eqnarray}      
where $C_B=2/27$ is the color factor, and $x_{1,2,3}$($y_{1,2,3}$) are the momentum fraction of $s(\bar s)$, $d(\bar d)$ and $u(\bar u)$, respectively.  A momentum fraction with a bar denotes $\bar x=1-x$.  We have used the property of Eq.~(\ref{PRLCDA}) and charge conjugation symmetry  to simplify our results of the perturbative coefficient functions.

We turn to the contributions from quark CEDM's. A typical diagram involving $\tilde d_s$ is given in Fig.~\ref{FF1}b. Again, because of the color structure the contributions involving the three-gluon vertex and the two-gluon part in the dipole interactions in Eq.~(\ref{LEDMs}) are zero.  Due to the helicity conservation, 
the CEDM of gluons or the CP-violating Weinberg operator~\cite{Wein} will not contribute at leading twist. For each CEDM $\tilde d_q(q=u,d,s)$ there are 56 diagrams. Calculating these diagrams, we have the result from $\tilde d_q$:
\begin{eqnarray} 
  d_\Lambda(Q) \biggr\vert_{\rm CEDMs} &=& e g_s  C_B \frac{4\pi \alpha_s}{Q^4}  \int [d^3x] [d^3y]  \biggr \{ \tilde d_s   \tilde{\mathcal H}_s ([x], [y]) \biggr [  {\mathcal V}_\Lambda ([x] ) {\mathcal V}_{\bar\Lambda} ([y]) - {\mathcal A}_\Lambda ([x] ) {\mathcal A}_{\bar\Lambda} ([y]) \biggr ]
\nonumber\\
     && + \left ({\mathcal V}_{\Lambda} ([x]) + {\mathcal A}_{\Lambda} ([x]) \right ) {\mathcal T}_{\bar\Lambda} ([y]) 
          \biggr [ \tilde d_d \tilde {\mathcal H}_d ([x],[y]) + \tilde d_u \tilde{\mathcal H}_u ([x],[y]) \biggr ]    \biggr \}~, 
 \label{DCEDMs} 
\end{eqnarray} 
with the perturbative coefficient functions:
\begin{eqnarray}
     \tilde{\mathcal H}_s ([x],[y] ) &=&Q_s \biggr [  \frac{1}{ \bar x_1\bar y_1} \left ( \frac{1}{x_3 y_3} + \frac{1}{x_2 y_2} \right ) 
         + \frac{2}{x_2 x_3 y_2 y_3} \biggr ] + Q_d \biggr [ \frac{1}{ x_1 y_1} \left ( \frac{1}{x_3 y_3} +\frac{1}{\bar x_2 \bar y_2} \right ) 
              + \frac{2}{x_3  y_3 \bar x_2 \bar y_2} \biggr ]
\nonumber\\
      && + Q_u \biggr [  \frac{1}{ x_1 y_1} \left ( \frac{1}{\bar x_3 \bar y_3} +\frac{1}{ x_2 y_2} \right ) +\frac{2}{x_2 y_2 \bar x_3 \bar y_3} \biggr ]~, 
\nonumber\\
 \tilde{\mathcal H}_d ([x],[y] ) &=&- Q_s \biggr [ \frac{2}{ x_2 y_2} \left (\frac{1}{x_3 y_3 } +\frac{1}{\bar x_1 \bar y_1} \right ) 
     +\frac{4}{x_3 y_3 \bar x_1 \bar y_1} \biggr ] - Q_d \biggr [\frac{2}{\bar x_2 \bar y_2} \left ( \frac{1}{x_3 y_3} + \frac{1}{x_1 y_1}\right ) 
         +\frac{4}{x_1 x_3 y_1 y_3} \biggr ] 
 \nonumber\\        
       &&  - Q_u \biggr [  \frac{2}{ x_2 y_2} \left ( \frac{1}{\bar x_3\bar y_3} +\frac{1}{x_1 y_1}\right ) 
              +\frac{4}{x_1 y_1 \bar x_3 \bar y_3} \biggr ]~,  
\end{eqnarray} 
where $Q_q (q=u,d,s)$ is the electric charge of the quark in unit of the positron charge $e$.  The function $\tilde {\mathcal H}_u$ is obtained 
from $\tilde {\mathcal H}_d$ by the exchange $Q_u \leftrightarrow Q_d$. From our results, $d_\Lambda(Q)$ scales like $Q^{-4}$ 
as expected. Furthermore, the factorization structures in Eqs.~(\ref{DEDMs},\ref{DCEDMs}) can be simply verified by using the helicity basis of $\Lambda$ in Eq.~(\ref{T3S}) and that of $\bar\Lambda$.

The factorization formulas summarized in Eqs.~(\ref{DEDMs},\ref{DCEDMs}) are the main results of our work. The total contribution from all quark CP violating dipoles to $d_\Lambda$ is the sum:
\begin{equation} 
   d_\Lambda (Q) = d_\Lambda(Q)\biggr\vert_{\rm EDMs} + d_\Lambda(Q) \biggr\vert_{\rm CEDMs}~. 
   \label{eq:sum}
\end{equation} 
With these analytical results, we can perform a numerical estimation on $d_\Lambda (Q)$ at $Q=M_{J/\psi}$ and infer the constrains on the quark EDMs and CEDMs using the BESIII result~\cite{BESIII:2025vxm}. However, it should be kept in mind  that to our formulas there are corrections from higher orders of $\alpha_s(Q)$ and from higher-twist contributions suppressed by $1/Q$.  These corrections could be substantial at $Q=M_{J/\psi}$ because $Q$  is not very large. These corrections may need to be further studied and are neglected here.

The numerical analysis requires non-perturbative input from the $\Lambda$ LCDAs, which have been determined using QCD sum rules~\cite{COZ0, COZ,Liu:2008yg,Liu:2014uha} and Lattice QCD~\cite{Bali:2015ykx,RQCD:2019osh,Bali:2024oxg,LatticeParton:2024vck,LPC:2025jvd}. The simplest choice is the asymptotic forms, given as~\cite{COZ0}
 \begin{equation} 
    {\mathcal A}_\Lambda ([x]) =f_\Lambda  f_{\rm as} ([x])=  120 f_\Lambda x_1 x_2 x_3, \quad  {\mathcal V}_\Lambda ([x])= {\mathcal T}_\Lambda ([x])=0~, 
 \end{equation} 
 with $f_\Lambda \approx 6\times 10^{-3}~{\rm GeV}^2$. Interestingly, we find that with the asymptotic forms, $d_\Lambda$ only receives contributions from the $s$-quark.  We have:
 \begin{equation} 
    d_\Lambda =  1.72 \times 10^{-4} d_s + 2.06\times 10^{-5} e\tilde d_s~. 
    \label{eq:num_asy}
\end{equation} 
However, it is inconsistent to use the asymptotic forms for finite $Q$, because these forms correspond to the scale in the limit $\mu\sim Q \to \infty$. 
It is noted that the asymptotic forms arise as the first term of conformal partial wave expansion of LCDAs in terms of orthogonal polynomials ${\mathcal P}_{nk}$~\cite{Braun:2008ia}. In the following, we include the next term in this expansion.

With the next term, the $\Lambda$ LCDAs takes the form~\cite{Bali:2015ykx}:
\begin{eqnarray} 
{\mathcal V}_\Lambda ([x], \mu) &=& -\frac{7}{2} \sqrt{\frac{3}{2}} f_{\rm as} ([x]) (x_3-x_2) [ \varphi_{10}(\mu) - 3\varphi_{11}(\mu) ]~, 
\nonumber\\
{\mathcal A}_\Lambda ([x],\mu) &=& \frac{1}{2} \sqrt{\frac{3}{2}}  f_{\rm as} ([x])  [-2 \varphi_{00}(\mu) +7 (x_3+x_2-2x_1) (\varphi_{10}(\mu) +\varphi_{11}(\mu) ) ]~, 
\nonumber\\
{\mathcal  T}_{\Lambda} ([x], \mu) &=& 7 \sqrt{\frac{3}{2}}  f_{\rm as} ([x])  (x_3-x_2) \pi_{10}(\mu)~, 
\end{eqnarray} 
where $\varphi_{00},\varphi_{10},\varphi_{11}, \pi_{10}$ are known as the shape parameters, with their one-loop $\mu$-dependence given in~\cite{Bali:2015ykx}. We first adopt the shape parameters extracted from QCD sum rules in the COZ analysis~\cite{COZ}. At $\mu_0=2{\rm GeV}$, they read~\cite{Bali:2015ykx}: 
\begin{equation} 
\phi_{00}(\mu_0) = 4.69\times 10^{-3}, \quad \phi_{10}(\mu_0)=1.39\times 10^{-3}~, \quad
\phi_{11}(\mu_0) = 1.05 \times 10^{-3}, \quad \pi_{10}(\mu_0) =1.32\times 10^{-3}~, 
\end{equation} 
in units of ${\rm GeV}^2$. Using these values and evolving to $\mu=Q=M_{J/\psi}$, we obtain
\begin{equation} 
    d_\Lambda =  1.94\times 10^{-3} d_s +1.47\times 10^{-4} (d_u+d_d)  +1.78\times 10^{-4} e\tilde d_s +7.39\times 10^{-5} e\tilde d_d 
       -9.53\times 10^{-5} e\tilde d_u~.  
       \label{eq:num_COZ}
\end{equation} 
We also consider the shape parameters determined from Lattice QCD~\cite{Bali:2024oxg,RQCD:2019osh}, obtained at $\mu_0=2~{\rm GeV}$~\cite{Bali:2024oxg}:
\begin{eqnarray} 
\phi_{00}(\mu_0) = 4.75\times 10^{-3}, \quad \phi_{10}(\mu_0)=0.563\times 10^{-3}~, 
\phi_{11}(\mu_0) = 0.242\times 10^{-3}, \quad \pi_{10}(\mu_0) =0.237\times 10^{-3}~, 
\end{eqnarray} 
again in ${\rm GeV}^2$. Using these lattice results at $\mu=Q=M_{J/\psi}$, we find
\begin{equation}
d_\Lambda = 5.29\times 10^{-4} d_s + 4.61\times 10^{-5}(d_u+d_d) + 6.21\times 10^{-5} e\tilde d_s+
1.98\times 10^{-5} e\tilde d_d - 2.14\times 10^{-5} e\tilde d_u~.
\label{eq:num_LQCD}
\end{equation}
Comparing these results with Eq.~(\ref{eq:num_asy}), one sees that the contributions from EDMs and CEDMs of $u$- and $d$-quark become nonvanishing once higher terms in the conformal expansion of the LCDAs are included. In our analysis, corrections from higher orders in $\alpha_s$ and from higher-twist contributions suppressed by powers of $1/Q$ are neglected. The main uncertainty arises from the input of the $\Lambda$ LCDAs, whose full functional forms are not yet precisely known. For the following analysis we will adopt the lattice-based input, as they are derived from first-principle calculations. While traditional lattice methods have accurately determined the first few moments of the $\Lambda$ LCDAs~\cite{Bali:2024oxg,RQCD:2019osh}, recent advances using large-momentum effective theory~\cite{Ji,JiC} enable access to the LCDAs without relying on moment expansions. In fact, one of the three $\Lambda$ LCDAs has already been investigated in this framework~\cite{LatticeParton:2024vck,LPC:2025jvd}.

      


\begin{figure}[t!!]
  \centering
  \begin{subfigure}{0.45\linewidth}
    \includegraphics[width=\linewidth]{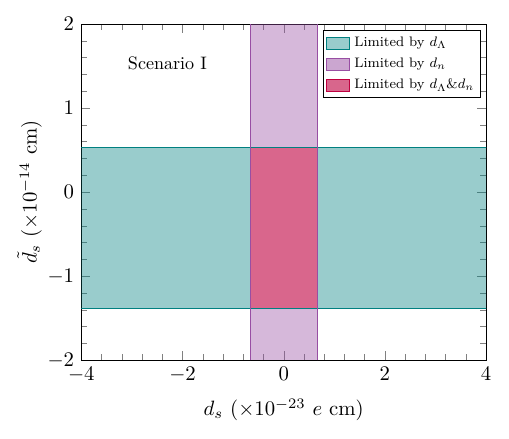}
    \caption{}
    \label{fig:bound1}
  \end{subfigure}
  \hspace*{0.3cm}
  \begin{subfigure}{0.45\linewidth}
    \includegraphics[width=\linewidth]{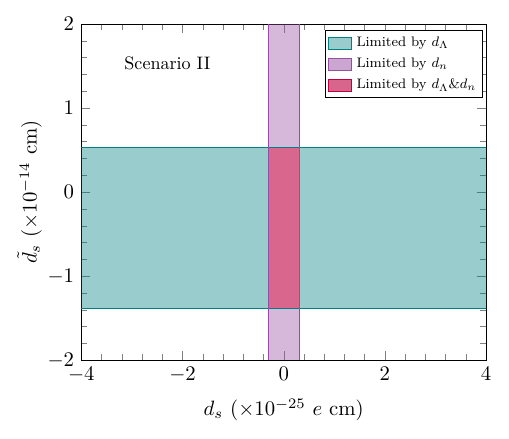}
    \caption{}
    \label{fig:bound2}
  \end{subfigure}

  \captionsetup{width=0.9\linewidth, font=small, labelfont=small}
  \caption{Constraints on the $s$-quark EDM $d_s$ and CEDM $\tilde d_s$ from the measurements on the $\Lambda$ and $n$ EDMs. (a) Scenario I: $d_s \gg d_u, d_d$ ; (b) Scenario II: $d_s=d_u=d_d$.
  }
  \label{Fig_bound}
\end{figure}

It is constructive to compare our results for $\Lambda$ with those for the neutron EDM $d_n$ induced by quark EDMs and CEDMs. The calculations based on QCD sum rules and Lattice QCD show that~\cite{Alarcon:2022ero}
\begin{equation} 
d_n = -(0.20 \pm 0.01) d_u + (0.78\pm 0.03) d_d + (0.0027\pm 0.0016) d_s  -(0.55\pm 0.28) e \tilde d_u -(1.1\pm 0.55) e \tilde d_d~,
\label{Eq:dn} 
\end{equation} 
where we have neglected the QCD $\theta$-term. Here, dipole couplings contribute to $d_n$ at $Q=0$ much more strongly than to $d_\Lambda$ at $Q=M_{J/\psi}$, and there is no contribution from $\tilde d_s$. If the dipole couplings are assumed to be of the same order, the strange-quark EDM $d_s$ makes a negligible contribution to $d_n$. In contrast, our numerical results given in Eqs.~(\ref{eq:num_COZ},\ref{eq:num_LQCD}) show that not only the $d_s$ contribution to $d_\Lambda$ is dominant under the same assumption, but also $\tilde d_s$ yields a nonzero contribution. This indicates that the upper bound of  $\Lambda$ EDM can provide a unique channel to constrain the strange-quark CEDM $\tilde d_s$.

To demonstrate this, in Fig.~\ref{Fig_bound} we present the constraints on the strange-quark EDM $d_s$ and CEDM $\tilde d_s$ derived from the current experimental bounds on the $\Lambda$ EDM $d_\Lambda$~\cite{BESIII:2025vxm} and the neutron EDM $d_n$~\cite{Abel:2020pzs}. To extract the limits on $d_s$ and $\tilde d_s$ from $d_\Lambda$ in Eq.(\ref{DL}), we employ our perturbative prediction in Eq.~(\ref{eq:num_LQCD}). For the neutron EDM, we use the central values from Eq.~(\ref{Eq:dn}) together with the experimental upper limit $\lvert d_n \rvert < 1.8 \times 10^{-26}~e~\mathrm{cm}$ reported in~\cite{Abel:2020pzs}. As an estimate, we presumably can neglect  the contributions from CEDM of $u$ and $d$.  For the quark EDMs, we consider two scenarios: (i) a hierarchy with $d_s \gg d_u, d_d$, motivated by the generic scaling $d_q \propto m_q$ in many new-physics models~\cite{Chupp:2017rkp}; and (ii) a $SU(3)$ flavor-symmetric model with $d_s=d_u=d_d$. In both scenarios, the $d_u$- and $d_d$- contributions  to $d_\Lambda$ are negligible, yielding nearly identical limits on $d_s$ and $\tilde d_s$. The only differences arise from the additional constraints imposed by $d_n$. The resulting bounds in two scenarios are displayed in Fig.~\ref{fig:bound1} and Fig.~\ref{fig:bound2}, respectively.

We observe that the $\Lambda$ EDM bound is particularly sensitive to $s$-quark CEDM $\tilde{d}_s$, while the neutron EDM bound provides a more stringent constraint on $s$-quark EDM $d_s$, excluding complementary regions of parameter space. Interestingly, the $\Lambda$ EDM bound alone does not restrict $\tilde{d}_s$ to a finite range: when $d_s$ and $\tilde{d}_s$ are expressed in the same scale units, the constraint would appear as an elongated band in the $(d_s, \tilde{d}_s)$ plane, reflecting the strong linear correlation between the two couplings. This correlation is substantially reduced once the neutron EDM input is included. Since $d_n$ in Eq.~(\ref{Eq:dn}) does not depend on $\tilde d_s$, and its experimental upper limit is much tighter than that of $d_{\Lambda}$, it provides the dominant restriction on $d_s$. Combining this with the $\Lambda$ EDM constraint, we obtain the following bound on the $s$-quark CEDM,
\begin{align}
  |\tilde{d}_s| \lesssim 1.4 \times 10^{-14}~{\rm cm}~.
\end{align}
 Importantly, the sensitivity to $\tilde{d}_s$ revealed here underscores the unique role of the $\Lambda$ EDM in probing strange-quark CP-violating CEDM, which remains inaccessible to the neutron EDM. This feature persists even without the assumptions adopted in our analysis.


 Before giving our summary, a discussion of the upper bound in  Eq.~(\ref{EQ1}) reported by BESIII~\cite{BESIII:2025vxm} may be useful for further studies. The contribution from $d_\Lambda$ to the decay $J/\psi \to \Lambda\bar\Lambda$ is through $J/\psi\to \gamma^* \to \Lambda\bar\Lambda$~\cite{HMM}. To the conversion of $J/\psi \to \gamma^*$, the EDM $d_c$ of charm quark may also contribute. However, if $J/\psi$ is a parity-eigenstate, or parity is conserved in the formation of $J/\psi$ from $e^+e^- \to J/\psi$, i.e., one neglects the weak interaction and those CEDMs in Eq.~(\ref{LEDMs}), the contribution vanishes. When parity-violating interactions are included, the effect is expected to be negligibly small. In addition, CP-violating effects may also arise from the strong-interaction decay $J/\psi \to \Lambda \bar\Lambda$ through the CEDMs of light quarks. Since the experimental limit applies to the sum of CP-violating effects, assuming this contribution from CEDMs in the decay is either negligible or not cancelled by $d_\Lambda$ partially or entirely, the upper bound on $d_\Lambda$ has been derived in Eq.~(\ref{EQ1}). These CEDM contributions could be estimated if one applies the collinear factorization of QCD to the decay.

 Finally, let us close this letter with a summary. Motivated by the recent BESIII measurement~($Q\gg \Lambda_{\text{QCD}}$)~\cite{BESIII:2025vxm}, we have performed the first perturbative QCD analysis of the $\Lambda$ EDM form factor $d_{\Lambda}(Q)$, tracing its origin to CP-violating quark dipole interactions. We have established a QCD factorization formula that expresses the $d_{\Lambda}(Q)$ form factor with the convolution of hard scattering coefficients and the $\Lambda$ LCDAs. With the input of LCDAs, we predict a numerical relation between the $\Lambda$ EDM and the quark EDMs and CEDMs. This relation enables direct constraints on these CP-violating dipole couplings from current and future $\Lambda$ EDM measurements. Our numerical analysis shows that the $\Lambda$ EDM is particularly sensitive to the strange-quark CEDM, yielding information complementary to that inferred from the neutron EDM. The framework can be readily extended to other hyperon EDMs, where future measurements and combined analyses would provide robust constraints on CP-violating dipole interactions beyond the SM. It would be also interesting to compare such bounds with those expected from high-energy experiments at the future electron-ion colliders~\cite{Boughezal:2023ooo,Wen:2024cfu,Huang:2025ljp} and lepton colliders~\cite{Cao:2025qua,Wen:2024nff}.

\par\vskip20pt

\noindent {\bf{Acknowledgements}}~The work is supported by National Key R\&D Program of China No.~2024YFE01098
00\&801. K.B. Chen is supported by National Natural Science Foundation of China (Grant No.~12005122), Shandong Province Natural Science Foundation~(Grant No.~ZR2020QA082), and Youth Innovation Team Program of Higher Education Institutions in Shandong Province (Grant No.~2023KJ126). X.G. He is supported by Natural Science Foundation of China (Nos. 12090064, 12375088, W2441004). X.B. Tong would like to thank Yingsheng Huang and Hao-Lin Wang for useful discussions, and he is supported by the Research Council of Finland, the Centre of Excellence in Quark Matter and projects 338263 and 359902, and by the European Research Council~(ERC, grant agreements No.~ERC-2023-101123801 GlueSatLight and No. ERC-2018-ADG-835105 YoctoLHC). The content of this article does not reflect the official opinion of the European Union and responsibility for the information and views expressed therein lies entirely with the authors.
\par

\end{document}